\documentclass[a4paper,11pt]{article}
\pdfoutput=1 % if your are submitting a pdflatex (i.e. if you have
\usepackage{jheppub} % for details on the use of the package, please
\usepackage{bm,amsmath,amssymb,slashed,graphicx,%
            enumerate,alltt,xspace,multirow,xcolor,mathrsfs}
\usepackage{fancyvrb}
\usepackage{cancel}

\usepackage[normalem]{ulem}
\usepackage{subcaption}

\usepackage{changes}

\definechangesauthor[name=pn, color=red]{pn}

\RequirePackage{braket}
\RequirePackage{graphicx}

\RequirePackage[numbers,sort&compress]{natbib}
\makeatletter
\g@addto@macro\bfseries{\boldmath}
\makeatother

\definecolor{labelkey}{rgb}{0,0.5,0.0}

\usepackage{listings}
\definecolor{darkgreen}{rgb}{0,0.4,0}
\newcommand\mathd{\mathrm{d}}

\newcommand{\as}{\alpha_s}

\newcommand{\PythiaEight}{{\tt Pythia8 {}}}

\title{Fits of $\alpha_s$ from event-shapes in the three-jet region: extension to all energies}
\preprint{
  \begin{flushright}
     MPP-2025-9
  \end{flushright}
}

\author[a,b]{Paolo Nason,}
\author[b,c]{Giulia Zanderighi}

\emailAdd{paolo.nason@mib.infn.it}
\emailAdd{zanderi@mpp.mpg.de}

\affiliation[a]{Universit\`a di Milano-Bicocca and INFN, Sezione di
  Milano-Bicocca, Piazza della Scienza 3,20126 Milano, Italy}
\affiliation[b]{Max-Planck-Institut f\"ur Physik,
Boltzmannstr. 8, D-85748 Garching, Germany} \affiliation[c]{Physik-Department,
  Technische Universit\"at M\"unchen, James-Franck-Strasse 1, 85748
  Garching, Germany}

\date{Received: date / Accepted: \today}

\abstract{This work is an extension of a previous
  publication~\cite{Nason:2023asn} where we fitted the strong coupling
  $\as$ together with the non-perturbative parameter $\alpha_0$ from
  event-shape and jet-shape distributions using power corrections
  computed in the three-jet region.
  In ref.~\cite{Nason:2023asn} only ALEPH data at the $Z$-pole were
  used in the fit. Here, instead, we include a large data sample from
  various $e^+e^-$ experiments at energies ranging from 22 to 207
  GeV and revisited the treatment of theoretical uncertainties.
  We find that the inclusion of different energies, while not changing
  the central fit result considerably, helps to disentangle the
  dependence of perturbative and non-perturbative corrections.
  Our best fit result is $\as(M_Z) =
  0.1181 \substack{ +0.0002 \\ -0.0005} \substack{ +0.0018 \\ -0.0021}$, 
  where the
  first error includes experimental uncertianties and the second one
  includes uncertainties associated with scale variation, mass
  effects, fit limits, non-perturbative schemes and non-perturbative
  uncertainties.

}

\keywords{Perturbative QCD, QCD Phenomenology, electron-positron
  scattering}

\begin{document}

\maketitle

\tableofcontents 

\section{Introduction}
\label{sec:intro}

The determination of the strong coupling $\alpha_s$ from shape
variables in the three-jet region in $e^+e^-$-annihilation has always
been considered the simplest context where to perform such
measurement, since the observables are dominated by terms of
order $\alpha_s$. In spite of the long history of $e^+e^-$ QCD studies, this
determination is controversial at present. On one side, these analyses
show clearly the validity of QCD predictions. On the other, when
requiring higher theoretical precision by including higher order in
pertubation theory, different determinations do not seem consistent
with each other, and some show a clear discrepancy with the world
average (see e.g.\ the discussion in the 2024 Particle Data Group
review on QCD~\cite{ParticleDataGroup:2024cfk}).

Part of the problem is due to the fact that the most precise
measurements of shape variables are carried out at the $Z$ peak, where
the energy is not large enough for non-perturbative corrections to be
negligible. In some analyses these effects are estimated using Monte
Carlo models, while others use analytic approaches.
One pitfall of the all analytic models used in the past is that they
were obtained by extrapolating non-perturbative corrections calculated
in the two-jet region to the regime of three-jets.
This problem was addressed in
refs.~\cite{Luisoni:2020efy,Caola:2021kzt,Caola:2022vea} and~\cite{Nason:2023asn}
where ways were found to estimate the power corrections in the
three-jet region. In particular, in ref.~\cite{Nason:2023asn} a fit of
shape variables relying on the new calculation of power corrections
was performed, leading to $\alpha_s$ values that were larger than those
obtained using the previous method, and well compatible with the world
average.
Subsequent work on the subject~\cite{Benitez:2024nav} has to some
extend addressed the issues raised in
refs.\cite{Luisoni:2020efy,Caola:2021kzt,Caola:2022vea,Nason:2023asn}
by restricting the fit-range to the two-jet
region. Ref.~\cite{Bell:2023dqs} has instead addressed the question of
whether theoretical uncertainties in the di-jet factorized predictions
can have a sizeable impact on the $\alpha_s$ determination. The
resummation of power-suppressed but logarithmically enhanced terms has
recently been studies in ref.~\cite{Dasgupta:2024znl}.

A further concern, that was pointed out in ref.~\cite{Nason:2023asn}
is the impact of resummed predictions in regions dominated by three
well separated jets, where its validity is questionable. In spite of
the fact that procedures to switch off resummation effects outside
their region of validity are well established (e.g. through the use of
modified logarithms or shape functions), it is often seen in practice
that resummation effects lead to an increase of differential
distributions by an amount which does not vanish even close ot the
kinematic upper limit of the distributions, and instead amounts to a
nearly constant factor larger than one. Given the availabity of more precise
perturbative calculations, these resummed predictions can end up
outside the fixed-order uncertainty band in the region where it
should be most reliable. It is clear that larger theoretical
predictions will typically lead to smaller values of the strong
coupling, at times incompatible with the world
average~\cite{ParticleDataGroup:2024cfk}.

A last issue, first pointed out in ref.~\cite{Salam:2001bd}, and which
emerged in particular in the analyses of ref.~\cite{Nason:2023asn}, is
the intrinsic ambiguity in shape variable definitions due to the
finite values of hadron masses. In ref.~\cite{Salam:2001bd} three
alternative schemes were proposed that lead to different results in
shape variable distributions.
Perturbative calculations deal with massless partons, and as such are
not affected by such ambiguities, while the experimentally measured
distributions are. It is however possible to use Monte Carlo
generators to infer the shape variable distributions in these alternative
schemes, starting from the distributions measured by the experiments in
their standard scheme. This was found to be the largest source of
uncertainties in ref.~\cite{Nason:2023asn}, suggesting that all
determinations of the strong coupling from shape varibales should
include this uncertainty.
If one instead believes that non-perturbative hadron-mass effects can
be absored into a re-parametrization of power-suppressed effects, one
needs to verify that, when using these alternative schemes, all mass effects
are absorbed into the non-perturbative function and that the
$\alpha_s$ determimnations obtained are fully compatible. If this is
not the case, a further uncertainty should be added. 

The present work is a follow-up of ref.~\cite{Nason:2023asn}
where we fitted the thrust, the $C$-parameter and the three-jet
resolution variable $y_3$ in terms of the strong coupling constant
$\alpha_S(M_Z)$ and a non-perturbative parameter $\alpha_0$.  The work
of ref.~\cite{Nason:2023asn} was based upon two ingredients: the
fixed-order calculation of shape-variable distributions at order
$\as^3$~\cite{Gehrmann-DeRidder:2007foh,Gehrmann-DeRidder:2007vsv,Gehrmann-DeRidder:2008qsl,DelDuca:2016csb},
and a calculation of non-perturbative corrections based upon the
findings of ref.~\cite{Caola:2022vea}.
In ref.~\cite{Nason:2023asn} we used only ALEPH data taken on the $Z$
peak, while in the present work we use a much larger data-set with
energies spanning from 22 to 207 GeV, available from different
experiments, thus including a total of about 1400 data points.

A number of improvements have been implemented in our fit
procedure.
First, we now use a dynamical scale in the
perturbative calculations, obtained by computing at leading order the average value of
the transverse momentum of the perturbative gluon giving rise to the
third jet for each value of the
observable.
A second improvement regards the inclusion, in the previous fits~\cite{Nason:2023asn}, of a
theoretical error in the $\chi^2$ definition.
This led to very small $\chi^2$ values, probably due to the fact that
we did not include bin-by-bin theoretical error correlations.
In this paper, we do not include a theoretical error in the
$\chi^2$ definition, and obtain more reasonable values of $\chi^2$.
The theoretical error is instead included as usual by performing
suitable variations on the setup of the theory predictions.
Third, we developed a procedure to determine the fit-range in an
automated way which is valid for different observables and at
different energies.
Next, we developed an improved, adaptive procedure to find the minimum
$\chi^2$ point in the ($\as$-$\alpha_0$)-plane that is fast enough also when
using a large data set.

Since in the previous paper we considered a single energy, we found
considerable degeneracy between the value of the strong coupling and
the non-perturbative parameter, that was lifted in part by fitting
simultaneously the three different observables. The inclusion of
different energies lifts this degeneracy
even further, allowing sensible fits also
to single observables.

The paper is organised as follows. In Sec.~\ref{sec:descr} we present
our theoretical framework and detail all changes compared to the
previous setup.
Sec.~\ref{sec:AppNP} recalls results and formulae for the
non-perturbative corrections and discusses a few changes compared to
ref.~\cite{Nason:2023asn}.
In Sec.~\ref{sec:data} we illustrate the
dataset included in the fit. In Sec.~\ref{sec:impl} we explain our fit procedure.
In Sec.~\ref{sec:res} we present and discuss
in detail our new fit result. In Sec.~\ref{sec:masses} we discuss the
inclusion of the heavy-hemisphere mass and hemisphere mass difference in our
fits.
Our conclusions are presented in Sec.~\ref{sec:conclu}.
%In Appendix~\ref{sec:masses} we discuss in inclusion of heavy jet-mass
%and jet-mass difference in the fits.

\section{Theoretical framework and changes compared to previous methods}
\label{sec:descr}
The variables of choice for our main fit are the thrust $T$
(or $\tau=1-T$), the $C$-parameter $C$, and three-jet resolution,
$y_{23}$. As discussed in ref.~\cite{Nason:2023asn} we do not include in the main fit
  the heavy-jet mass $M_H^2$ and the jet-mass difference $M_D^2$, since for them the non-perturbative corrections
  have some undesirable features in the two-jet limit.
We do however discuss them  in Sec.~\ref{sec:masses}, and perform a fit including them,
showing that to some extent our calculation of power corrections describes them better than
the previous (2-jet based) one.

All the variables that we consider in this work
are defined in Sec.~2 of ref.~\cite{Nason:2023asn}, and we will not
repeat the definition here. Similarly, ambiguities related to hadron
masses are discussed in Sec.~3 of the same reference. Here we adopt
the same default scheme choice (i.e. the $E$-scheme) and consider the
$p$, $D$ and standard scheme as variations. As in the previous work,
heavy-quark mass-effects are corrected for using Monte Carlos (see
Sec.~6.2 of ref.~\cite{Nason:2023asn}).

\subsection{Calculation of the differential distributions}

Our perturbative description of a shape-variable distribution $v$
relies on the perturbative computation of $e^+e^-\to 3$ jets at NNLO,
performed using the public EERAD3
code~\cite{Gehrmann-DeRidder:2007foh,Gehrmann-DeRidder:2007vsv,Gehrmann-DeRidder:2008qsl},
which is based on the antenna subtraction
formalism~\cite{Gehrmann-DeRidder:2005btv}.

The starting point is the cumulative distribution at center-of-mass energy $Q$ and renormalization
scale $\mu_R$, which  can be written as
\begin{equation}
  \bar \Sigma_{\rm NNLO}(v,\mu_R,Q) = \int_v^1 dv'
 \left[\frac{1}{\sigma_{\rm \scriptscriptstyle NNLO}}\frac{d \sigma_{\rm \scriptscriptstyle NNLO}(v',\mu_R,Q)}{d v'}   \right]_{\rm trunc}\,,
\label{eq:SigmaNNLO}
\end{equation}
where the suffix ``trunc'' indicates that the ratio in the square
bracket is expanded and truncated at NNLO order (see Eq.(5.2) of
ref.~\cite{Nason:2023asn}).
We have
\begin{equation}
\bar \Sigma_{\rm NNLO}(v,\mu_R,Q) +\Sigma_{\rm NNLO}(v,\mu_R,Q) = 1\,, 
  \end{equation}  
where $\Sigma_{\rm NNLO}(v,\mu_R,Q)$ is given in Eq.~(5.1) of ref.~\cite{Nason:2023asn}. 
Using $\bar \Sigma_{\rm NNLO}$ rather than $\Sigma_{\rm NNLO}$ (as was done in
ref.~\cite{Nason:2023asn}), avoids the need to specify how to handle
the (divergent) $v\to 0$ region.

Here, at variance with ref.~\cite{Nason:2023asn}, we evaluate
eq.~\eqref{eq:SigmaNNLO} using a dynamical scale $\mu_R$, that is a
function of the integration limit $v$. To obtain the contribution to a
bin of the shape variable, we take the difference between $\bar
\Sigma(v,\mu_R(v),Q)$ evaluated at the endpoints of the bin. 

Our choice for the dynamical scale is the average transverse momentum
for a given value of $v$ relative to the thrust axis computed at Born
level.\footnote{This corresponds to $Q$ times the wide broadening.} It
is obtained by first histogramming the average transverse momentum as
a function of $v$ computed at Born level. The result is then fitted
with a simple parametrization of the form $\langle k_t\rangle = a\cdot
v^b$. This average transverse momentum is taken to be our central
scale choice $\mu_0$. As usual, we will consider variations by a
factor two above and below this value.

The second ingredient for the computation of the distributions is the
implementation of the non-perturbative corrections. We implemented a
few modifications with respect to ref.~\cite{Nason:2023asn}. In
particular we now separate variations associated with how the
non-perturbative shift is implemented (schemes a, b, and c) and with
the effect of estimated quadratic power corrections. We discuss these aspects in
detail in the following section. 

\subsection{Non-perturbative corrections schemes and uncertainties}
\label{sec:AppNP}
We compute the non-perturbative cross section to the cumulative
distributions of a shape variable $v$ using formulae (4.29) through
(4.32) of ref.~\cite{Nason:2023asn}, which make use of the function
$h_v(\eta,\phi)$, defined in eq.~(4.6) of the same
reference.\footnote{In eq.~(4.6) of ref.~\cite{Nason:2023asn} a factor Q is missing.}.
In practice, in our calculation, rather than using
the definition of $H_{\rm NP}$ given in Eq.~(4.32) we use the value
that can be inferred from Eq.~(4.29) and (5.5), that is commonly
adopted in dispersive models~\cite{Akhoury:1995sp,Dokshitzer:1995qm,Dokshitzer:1997iz,Dokshitzer:1998pt}.

In order to estimate quadratic power suppressed effects
we now define $h_v(\eta,\phi)$ in the following way:  
\begin{equation}
  {h}_v(\eta,\phi) =\lim_{l_\perp\to 0}\tilde{h}_v\left(\eta,\phi,\frac{|l_\perp|}{Q}\right)\,,
\end{equation}
with 
\begin{equation}
  \tilde{h}_v\left(\eta,\phi,\frac{|l_\perp|}{Q}\right) = \frac{Q}{|l_\perp|}\left[ v(\{P\},l) - v(\{p\}) \right]. 
\end{equation}
Furthermore, we introduce the following functions
\begin{equation}
  \tilde\zeta_1\left(v,\frac{|l_\perp|}{Q}\right) = \sum_{\rm dip} \frac{C_{\rm dip}}{C_F} \left(\frac{\mathd \sigma_B}{\mathd v}\right)^{-1}
  \int \mathd \sigma_B(\Phi_B) \int \mathd\eta \frac{\mathd\phi}{2\pi}
  \frac{ \theta\left(v(\{P\},l)-v\right)  -  \theta(v(\{p\})-v)}{|l_\perp|/Q}, 
\label{eq:tildezeta}
\end{equation}
and 
%By expanding the theta function in the above equation using eq.~\ref{eq:zetatildedef}  we find the alternative formula
\begin{eqnarray}
  \tilde\zeta_2\left(v,\frac{|l_\perp|}{Q}\right) = \sum_{\rm dip} \frac{C_{\rm dip}}{C_F} \left(\frac{\mathd \sigma_B}{\mathd v}\right)^{-1}
  \int \mathd \sigma_B(\Phi_B)\,\delta\left(v(\{p\})-v\right)
  \int \mathd\eta \frac{\mathd\phi}{2\pi}\tilde{h}_v\left(\eta,\phi,\frac{|l_\perp|}{Q}\right) \,,
\end{eqnarray}
where the second one is obtained from the first one
by Taylor expanding the first
$\theta$-function in eq.~\eqref{eq:tildezeta}.
Both expressions are modifications of the $\zeta(v)$ function that differ from it
  by terms of order $|l_\perp|$, and thus have the expansion
\begin{equation}\label{eq:c12def}
  \tilde\zeta_{1/2}\left(v,\frac{|l_\perp|}{Q}\right)) = \zeta(v) \left[1+ \frac{|l_\perp|}{Q} \cdot c_{1/2}\left(v,\frac{|l_\perp|}{Q}\right)  \right]\,, 
\end{equation}
% zeta(:,1) -> v 
% zeta(v) zeta(:,2) 
% c_1 corresponds to 2-zeta -> zeta(:,3) 
% c_2 corresponds to 3-zeta -> zeta(:,4)
where $c_{1/2}$ have a well defined limit for $|l_\perp| \to 0$. We
compute the functions $\tilde\zeta_{1/2}$ numerically, by setting up
  a generator for $q\bar{q} g$ production plus an extra radiation with a fixed
  transverse momentum $|l_\perp|$ in the radiating dipole rest frame,
  including all recoil effects, as described in Sec.~4.4 of ref.~\cite{Nason:2023asn}.
  We choose $|l_\perp|/{Q}= 0.01$, which we assume to be
small enough to be close to the $|l_\perp| \to 0$ limit. The functions
  $c_{1/2}$ are then extracted using eq.~(\ref{eq:c12def}).
We thus define
\begin{equation} \label{eq:Kdef}
  K_{1/2}(v)= C\cdot c_{1/2}(v)\cdot  H_{\rm NP},
%  K(v)= C\cdot \frac{ \tilde{\zeta}(v)- \zeta(v)}{\zeta(v)} \cdot \frac{Q_0}{\lambda_0} \cdot  H_{\rm NP},  
\end{equation}
where $C$ is a coefficient of order 1 that we take equal to 1,
and in our implementation of the non-perturbative correction we
use as default the shift
\begin{equation}
  \delta_{\rm NP}(v)= \zeta(v)\cdot H_{\rm NP} \cdot\left(1+ K_1(v)\right). 
\end{equation}
The cumulative distribution including non-perturbative corrections is then given by
\begin{equation}
  \Sigma^{\rm (a)}(v)=\Sigma_{\rm NNLO}(v-\delta_{\rm NP}(v)).
\end{equation}
This defines our default scheme for the calculation of non-perturbative corrections, which we denote as (a).
As alternatives we also propose schemes
(b) and (c) defined as
\begin{eqnarray}
  \Sigma^{\rm (b)}(v)&=&\Sigma_{\rm LO}(v-\delta_{\rm NP}(v)) + \Sigma_{\rm NNLO}(v) - \Sigma_{\rm LO}(v)\,, \\
  \Sigma^{\rm (c)}(v)&=&\Sigma_{\rm NNLO}(v) - \delta_{\rm NP}(v) \frac{\mathd\Sigma_{\rm LO}(v)}{\mathd v}.
\end{eqnarray}
In ref.~\cite{Nason:2023asn} we also considered a scheme (d), that is as scheme (a) but without including
the $K(v)$ term.
In the present work instead we prefer to include explicitly separate variations associated with missing higher
order power corrections in the $\zeta$ functions. We define for this purpose
\begin{equation}
  \delta \Sigma_{\rm NP}(v) = | \Sigma^{\rm (a)}(v) - \Sigma_{\rm NNLO}(v)|\cdot \max(|K_1(v)|,|K_2(v)|).
\end{equation}
%where $K'$ is defined as $K$ in formula ref.~\ref{eq:Kdef}, but using $\tilde{\zeta}'$ instead of  $\tilde{\zeta}$.
The variations labeled npup (npdn) in the results presented in this paper are obtained by adding (subtracting) $ \delta \Sigma_{\rm NP}(v) $
to the cumulants computed according to our default method~(a).

\subsection{Determination of fit ranges}\label{sec:fitrange0}

We first compute the lower limit of the fit range as the peak position
in the data multiplied by a factor $C_{ll}$, that we set by default equal to two. This choice guarantees that
one is sufficiently far away from the region where resummation effects
become important. The peak position is loosely fitted
with a simple parametrization of the form $v_{\rm peak}(E) = v_{0}\cdot (91.2/E)^a$, where
$v_{0}$ is the position of the peak at energy 91.2 GeV.
In the case of $y_3$ we freeze the lower limit that at the $Z$-pole
for energies above $M_Z$. The reason to do this is that at higher
energies the peak position reaches very low $y_3$ values. 
Furthermore, we also determine a lower limit by imposing that the running
renormalization scale never falls below 4~GeV. The final lower limit is taken equal
to the maximum of the limits found with the two methods.

The upper limits of the fit-ranges are taken close to the kinematical
bound for three parton production, and are set to 0.6 for the
$C$-parameter and 0.3 for $\tau$ and $y_3$.

\section{Description of data set used}\label{sec:dataset}
\label{sec:data}

To perform our fits, we used data collected at energies between 22
and 207 GeV from the ALEPH~\cite{ALEPH:2003obs},
DELPHI~\cite{DELPHI:1999vbd,DELPHI:2000uri,DELPHI:2003yqh},
JADE~\cite{MovillaFernandez:1997fr},
L3~\cite{L3:1992nwf,L3:2002oql,L3:2004cdh},
OPAL~\cite{OPAL:1992dnu,OPAL:1993pnw,OPAL:2004wof},
SLD~\cite{SLD:1994idb} and TRISTAN~\cite{TOPAZ:1993vyh} experiments,
as well as the combined analysis of $y_3$ using data from JADE and
OPAL~\cite{JADE:1999zar}, which we refer to as JADE-OPAL. The
available energies and the observables used, for each experiment, are
listed in Table~\ref{tab:data}. We have omitted data on $y_3$ by JADE
and OPAL, for energies that are already included in JADE-OPAL. With
this set of data, available in the HEPData
database~\cite{Maguire:2017ypu}, taking into account our fit range, we
include up to 915 data points in the fit.

\begin{table}[htb]
\begin{center}
\begin{tabular}{|l|l|l|}
\hline
  Obs. & Experiment & Energies  in GeV \\
\hline\hline
  $C$&ALEPH~\cite{ALEPH:2003obs}      &   91.2 133   161   172   183   189   200   206   \\
  \hline
 $C$&DELPHI~\cite{DELPHI:1999vbd,DELPHI:2000uri,DELPHI:2003yqh}     &   45   66   76  91.2 133   161   172   183   189   192   196   200   202   205   207    \\
  \hline
 $C$&JADE~\cite{MovillaFernandez:1997fr}       &   35  44 \\
  \hline
 $C$&L3~\cite{L3:1992nwf,L3:2002oql,L3:2004cdh}         &   91.2 130.1  136.1  161.3  172.3  182.8  188.6  194.4  200   \\
  \hline
 $C$&OPAL~\cite{OPAL:1992dnu,OPAL:1993pnw,OPAL:2004wof}       &   91.2 133  177  197 \\
  \hline
 $C$&SLD~\cite{SLD:1994idb}        &   91.2 \\
  \hline
 $T$&ALEPH~\cite{ALEPH:2003obs}      &   91.2 133  161  172  183  189  200  206   \\
  \hline
 $T$&DELPHI~\cite{DELPHI:1999vbd,DELPHI:2000uri,DELPHI:2003yqh}     &   45   66   76  91.2 133   161   172   183   189   192   196   200   202   205   207     \\
  \hline
 $T$&JADE~\cite{MovillaFernandez:1997fr}       &   35  44 \\
  \hline
  $T$&L3~\cite{L3:1992nwf,L3:2002oql,L3:2004cdh}         &   41.4   55.3   65.4   75.7   82.3   85.1   91.2 130.1   136.1   161.3   172.3     \\
    &         &     182.8   188.6   194.4   200  \\
  \hline
 $T$&OPAL~\cite{OPAL:1992dnu,OPAL:1993pnw,OPAL:2004wof}       &   91.2 133   177   197 \\
  \hline
 $T$&SLD~\cite{SLD:1994idb}        &   91.2 \\
  \hline
 $T$&TRISTAN~\cite{TOPAZ:1993vyh}    &   58 \\
  \hline
 $y_3$&ALEPH~\cite{ALEPH:2003obs}      &   91.2 133  161  172  183  189  200  206  \\
   \hline
 $y_3$&JADE~\cite{MovillaFernandez:1997fr}       &   22\\
   \hline
 $y_3$&JADE-OPAL~\cite{JADE:1999zar}   &   35 44 91.2 133 161 172 183 189 \\
   \hline
 $y_3$&OPAL~\cite{OPAL:1992dnu,OPAL:1993pnw,OPAL:2004wof}       &   177  197\\
   \hline
 $y_3$&TRISTAN~\cite{TOPAZ:1993vyh}    &   58\\
\hline 
\end{tabular}
\end{center}
\caption{\label{tab:data}
Summary of observables, experiments and energies used in the fit.}
\end{table}

%\section{Fit procedure, default setup and variations}

\section{Fit procedure}
\label{sec:impl}
We consider together all histogram bins of all shape variables from all
experiments and at all energies that we include in the fit and label them
with a single index.  The $\chi^2$ is then defined as
\begin{equation}
  \chi^2=\sum_{ij} \left(v_i-v_i^{(\rm th)}\right) V^{-1}_{ij} \left(v_j-v_j^{(\rm th)}\right), 
\end{equation}
where the covariance matrix is defined as\footnote{Note that, at
  variance with Eq.~(6.1) of ref.~\cite{Nason:2023asn}, no theoretical error is
  included here.}
\begin{equation}
  V_{i j}=\delta_{ij}R_i^2 + (1-\delta_{ij})C_{ij}R_iR_j + {\rm Cov}^{(\rm Sys)}_{ij} \,,
\end{equation}
where $R_i$ is the statistical error, $S_i$ the systematic error,
$C_{ij}$ the statistical correlation matrix, and ${\rm Cov}^{(\rm
  Sys)}_{ij}$ the covariance matrix for the systematic errors. When
systematic uncertainties are asymmetric we take the maximum of the
two. For the calculation of the statistical and systematic covariance
matrix we refer to ref.~\cite{Nason:2023asn}, Eq.~(6.2) and (6.3),
respectively.

In order to find the minimum $\chi^2$, as function of $\as$ and
$\alpha_0$, we compute the $\chi^2$ on a rectangular grid of $\as$ and
$\alpha_0$ points, and find the minimum value on the grid. The true minimum
value of $\as$ ($\alpha_0$) is estimated using a quadratic interpolation
of the three points on the grid: the grid minimum-point and its two neighboring
points in the $\as$ ($\alpha_0$) direction. The $\chi^2$ value is
taken as the average of the two interpolated values. While in
ref.~\cite{Nason:2023asn} we could easily use grids with a large
number of points to obtain an accurate result with this method, this is
no longer feasible in the present case because of the very large
number of data points.
Therefore, we start the procedure with a grid with a modest number of
points, find a minimum and iterate the procedure taking the minimum as
the new center of the grid, reducing the grid range while keeping the
same number of grid points.
In practice, we typically start with a grid centered around $\as=0.12$
and $\alpha_0=0.5$, we use a grids which has 11 by 11 points, such
that the maximum variation in $\as$ is $|\Delta \as| = 0.01$ and
$|\Delta \alpha_0|=0.5$. At each iteration we half the maximum
variation. In general we perform four iterations. We have checked that
performing more iterations the result remains stable. Once we have
found the minimum, we compute $\chi^2$ values around the minimum on a
relatively fine but sufficiently large grid, in order to compute the $\Delta
\chi^2$ contours.

\section{Results}
\label{sec:res}

\begin{figure}[htb]
  \begin{center}
    \includegraphics[width=0.6\linewidth, page=1]{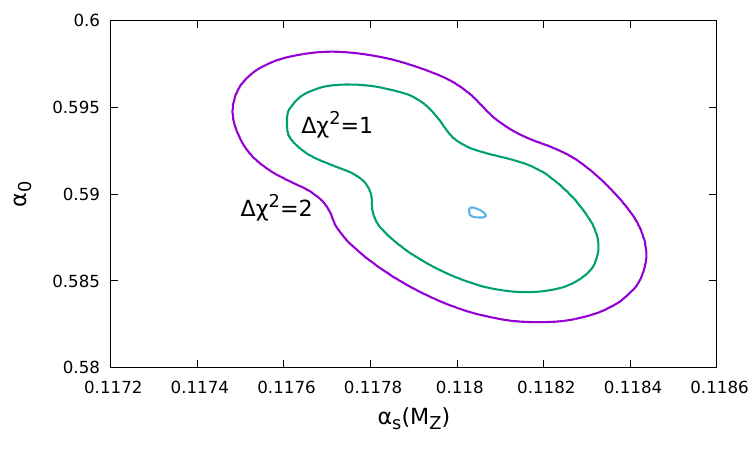}
    \caption{Fit central point and constant $\Delta\chi^2$ contours, for the simultaneous fit using $C$, $\tau$ and $y_3$. }
    \label{fig:CTy3-all}
  \end{center}
\end{figure}

We begin by presenting the simultaneous fit of the three shape
variables $T$, $C$ and $y_3$, which we refer to as the CTy3 fit, using the 
full data set described in Sec.~\ref{sec:dataset}. 
We use our default central setup.  The minimum $\chi^2$
and the contours for one and two units of $\chi^2$ above the minimum
are shown in fig.~\ref{fig:CTy3-all}.
This fit includes $N_{\rm deg} = 895$ bins and
has $\chi^2/N_{\rm deg} = 1.26$. The central value of $\alpha_s$ is in good
agreement with the world average and the value of the non-perturbative
parameter agrees well with previous determinations (see e.g. \cite{Dasgupta:2003iq}).

\begin{figure}[htb]
  \begin{center}
    \includegraphics[width=0.49\linewidth, page=1]{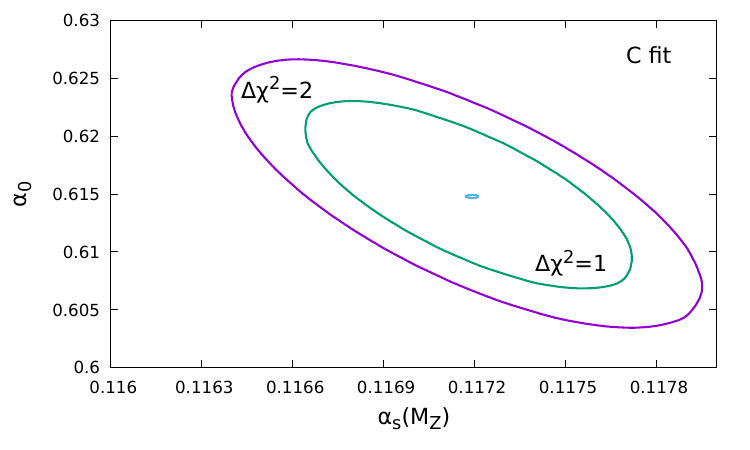}
    \includegraphics[width=0.49\linewidth, page=1]{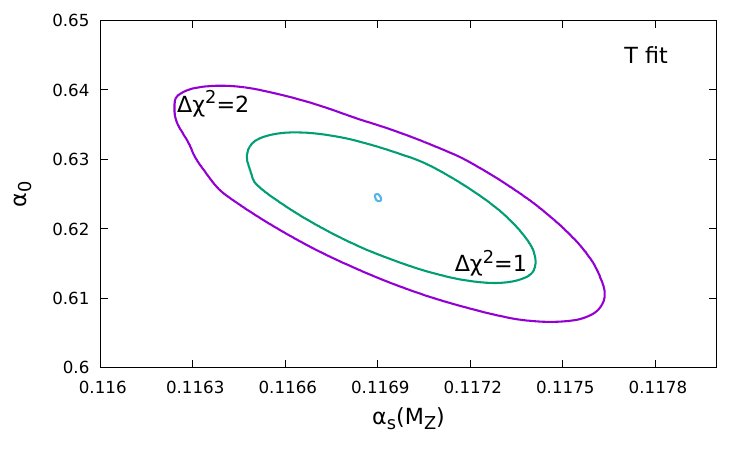}
    \includegraphics[width=0.49\linewidth, page=1]{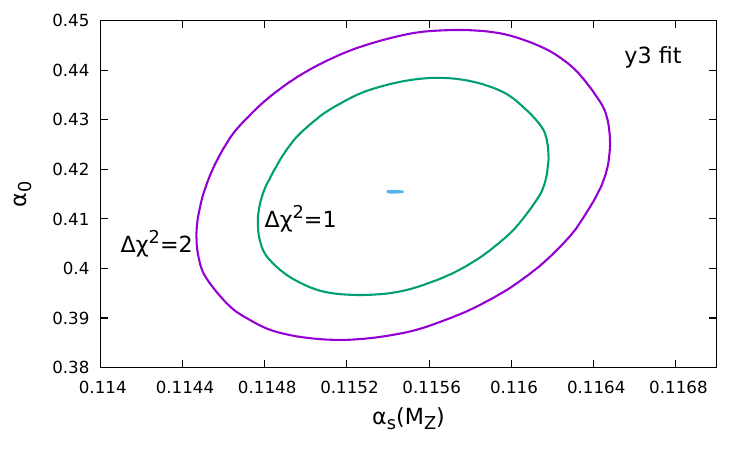}
    \includegraphics[width=0.49\linewidth, page=1]{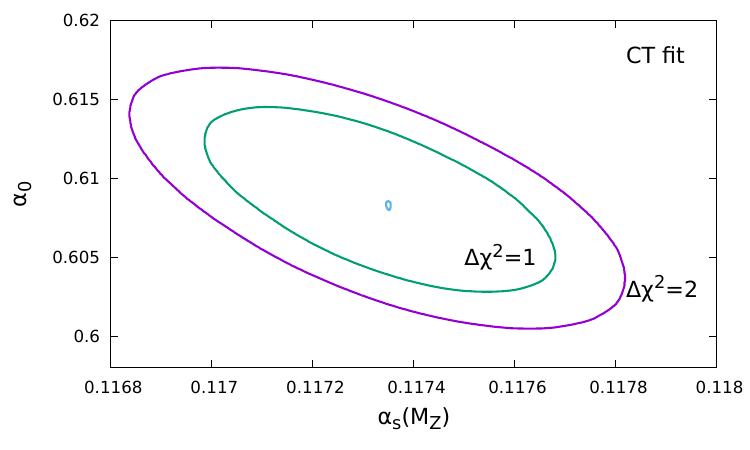}
    \caption{Fit central point and Constant $\Delta\chi^2$ contours for the fit of the $C$-parameter (left top) , $\tau$ (right top),
     y3 (left bottom) and CT (right bottom)}
    \label{fig:C-T-y3-CT}
  \end{center}
\end{figure}

\begin{figure}[htb]
  \begin{center}
    \includegraphics[width=0.7\linewidth, page=1]{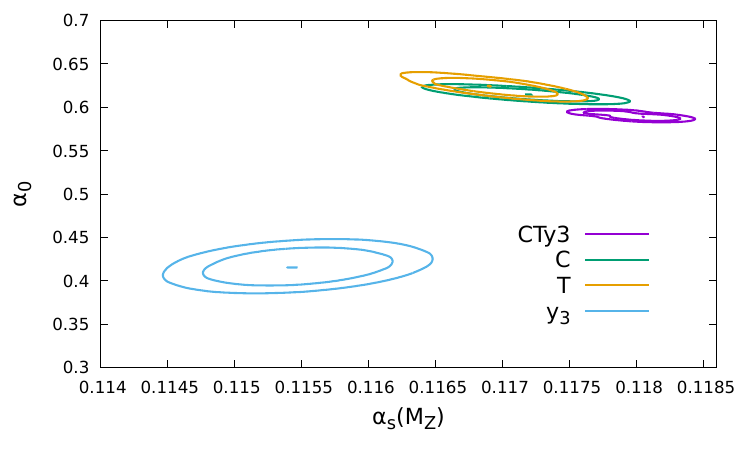}
    \caption{Fit central point and Constant $\Delta\chi^2$ contours for the CTy3, C, T and y3 fits reported on the same plot for comparison.}
    \label{fig:allPlots}
  \end{center}
\end{figure}

The result of the fits performed using each observable separately and using the $C$-parameter and $\tau$ simultaneously
are presented in Figs.~\ref{fig:C-T-y3-CT}, 
where C, T, y3 and CT label the
$C$-parameter, $\tau$, $y_3$ and the simultaneous fits of $C$ and $\tau$, respectively.
In fig.~\ref{fig:allPlots}
we report the CTy3, C, T and y3 fit on the same plot for comparison.

In table~\ref{tab:allchi2}
\begin{table}[htb]
  \begin{center}
    \begin{tabular}{|l|c|c|c|c|c|}
      \hline
      Fit name  & d.o.f.  &  $\chi^2$ & $\chi^2/{\rm d.o.f.}$   & $\as(M_Z)$   &  $\alpha_0$   \\
      \hline
      CTy3  & 895  &  1125.6   & 1.26                 &   0.1181    &   0.5902      \\
      \hline
      CT    & 744  &  952.6    & 1.28                 &   0.1173    &   0.6089      \\
      \hline
       C    & 292  & 269.0     & 0.921                &    0.1172   &   0.6148      \\
      \hline
       T    & 452  & 651.7     & 1.44                 &    0.1169   &   0.6245      \\
      \hline
       y3   & 151  & 71.6      & 0.47                 &    0.1155   &   0.4151      \\
      \hline
    \end{tabular}
    \caption{\label{tab:allchi2} Results for all fits.}
  \end{center}
\end{table}
we summarise the results of all fits. In table \ref{tab:indchi2}
\begin{table}[htb]
  \begin{center}
    \begin{tabular}{|l|c|c|c|}
      \hline
      Obs.   & d.o.f.  &  $\chi^2$ & $\chi^2/{\rm d.o.f.}$ \\
      \hline
       $C$   & 292  & 278.2     & 0.95    \\
      \hline
      $\tau$ & 452  & 659.7     & 1.465   \\
      \hline
       $y_3$ & 151  & 132.7     & 0.879   \\
      \hline
    \end{tabular}
    \caption{\label{tab:indchi2} Individual $\chi^2$ contribution
      of each observable to the combined CTy3fit.}
  \end{center}
\end{table}
we present the $\chi^2$ and $\chi^2/{\rm d.o.f.}$ values computed
for each separate observable at the $\as$ and $\alpha_0$ values
determined with the CTy3 fit.\footnote{The sum of these $\chi^2$ values does
not equal the total $\chi^2$ of the CTy3 fit, since correlations among
different observables are not included there.}

We observe that the $\chi^2/{\rm d.o.f.}$ value are quite acceptable for all fits.
The T fit yields the largest  $\chi^2/{\rm d.o.f.}$, and the y3 fit the smallest, both
as individual fits and as contributions to the CTy3 fit.
The individual fit to $y_3$ leads to a $\chi^2/{\rm d.o.f.}$
value that is very small (0.47), and to an $\alpha_0$ value
that is not consistent with the other two determinations. However, due to the very low
$\chi^2/{\rm d.o.f.}$ value, the $\chi^2$ for $y_3$ obtained using the $\alpha_0$ and $\as$
values of the CTy3 is still very reasonable.

Notice that the $\as(M_Z)$ result of the CTy3 fit is larger than all
individual determinations. This is not surprising if we look at the correlation between
$\as(M_Z)$ and $\alpha_0$ in fig.~\ref{fig:C-T-y3-CT}. In order to accommodate all observables with a single fit the
value of $\alpha_0$ should be higher than the preferred $y_3$ value, and lower for the
C and T fits, which implies larger values of $\alpha(M_Z)$ in both cases.
\subsection{Scale variations}
Next, we consider the impact of performing a renormalizaton scale
variation by factor 2 up and down in the theory predictions. The
result is shown in Table~\ref{tab:scale}.\footnote{Note that the number of bins depends slightly on the renormalization
scales. This, as explained in Sec.~\ref{sec:fitrange} is due to the
fact that the lower limit is adjusted in such a way as not to include
bins where the running renormalization scale falls belos 4~GeV.}
\begin{table}[htb]
\begin{center}
  \begin{tabular}{|c|c|c|c|c|c|}
    \hline
    Observable & scale & $\alpha_s(M_Z)$ & $\alpha_0$ & $N_{\rm bins}$ &  $\chi^2/N_{\rm deg}$ \\
  \hline 
     & $\mu_0/2$       & 0.1167 & 0.67  &  816 & 2.38\\
CTy3 & $\mu_0$         & 0.1181 & 0.59 &  895 & 1.26\\
     & $2 \mu_0$       & 0.1167 & 0.58  &  915 & 1.60\\
  \hline 
     & $\mu_0/2$       & 0.1146 & 0.75 &  705 & 1.38\\
CT   & $\mu_0$         & 0.1175 & 0.61 &  744 & 1.28\\
     & $2 \mu_0$       & 0.1211 & 0.50 &  744 & 1.34\\
  \hline
     & $\mu_0/2$       & 0.1141 & 0.76 &  285 & 0.99\\
  C  & $\mu_0$         & 0.1169 & 0.62 &  292 & 0.92\\
     & $2 \mu_0$      & 0.1212 & 0.51 &  292 & 0.95\\
\hline
     &$\mu_0/2$       & 0.1159 & 0.71 &  420 & 1.52\\
  T  &$\mu_0$         & 0.1168 & 0.63 &  452 & 1.44\\
     &$2 \mu_0$       & 0.1208 & 0.53 &  452 & 1.52\\
\hline
      & $\mu_0/2$       & 0.1122 & 0.03 &  111 & 0.53\\
 y3   & $\mu_0$         & 0.1155 & 0.42 &  151 & 0.47\\
      & $2 \mu_0$       & 0.1157 & 0.52 &  171 & 1.61\\
\hline
  \end{tabular}
\end{center}
\caption{Impact of scale variations on the results for all fits.}
\label{tab:scale}
\end{table}
We see that in the CTy3 fit the value of $\alpha_s$ decreases slightly (by about
1.2\%) both when varying the scale up or down. We find, however, larger
values of $\chi^2$, especially when using a
scale $\mu_0/2$. This result is better understood by examining the
effect of scale variations in the individual C, T and y3 fits.
In these cases we obtain much more reasonable $\chi^2$ values,
although always worse than the central value.
However the $\alpha(M_Z)$ and $\alpha_0$ value obtained in the y3 fit for the scale choice $\mu_0/2$ is not
compatible with that of the other determinations, which leads to the large
$\chi^2/{\rm d.o.f.}$ value in the CTy3 fit.
Some discrepancies of the fitted parameters in the C and T fits
alone is also observed, but is not quite as strong, so that
in the CT fit the increase in $\chi^2$ due to scale variations
is not so large.

In conclusion, if we consider fits using individual variables, or the
combined CT fit, we find large scale variations, of the order of
2-3\%, while in the combined CTy3 fit the scale uncertainty amounts
to about 1\% on $\alpha_ss(M_Z)$.
This is in part due to the correlation between $\alpha_s(M_Z)$ and
$\alpha_0$. In fact, when the scale increases $\alpha_s(M_Z)$
increases and $\alpha_0$ decreases for the C, T and CT fits, while for
the y3 fit both $\alpha_s(M_Z)$ and $\alpha_0$ increase.

\subsection{Variations of other theory setting}
When considering only scale variations, the extracted value of $\as(M_Z)$ is confined in a range of 1.2\%,
suggesting that perhaps an extraction of $\as$ with a 0.6\%{} accuracy may be possible. Unfortunately, by
considering the other sources of uncertainties that were studied in ref.~\cite{Nason:2023asn}, this does not seem to be
the case. All the considered uncertainties for the CTy3 fit are reported in table~\ref{tab:CTy3var}.

\begin{table}[htb]
  \begin{center}
\begin{tabular}{|l|c|c|c|c|c|}
\hline
 Variation & $\as(M_Z)$ & $\alpha_0$ & $\chi^2 $ & $\chi^2/{\rm d.o.f.}$ & d.o.f. \\
\hline
default & {\bf 0.1181} & 0.5902 & 1125.6463 & 1.2577 & 895 \\
\hline
$\mu_R=\mu_0/2$ & 0.1167 & 0.6683 & 1940.0007 & 2.3775 & 816 \\
\hline
$\mu_R=2\mu_0$ & 0.1167 & 0.5846 & 1465.9393 & 1.6021 & 915 \\
\hline
std scheme & 0.1173 & 0.5347 & 1090.8732 & 1.2202 & 894 \\
\hline
p scheme & {\bf 0.1160} & 0.5624 & 1051.1005 & 1.1757 & 894 \\
\hline
D scheme & {\bf 0.1199} & 0.7252 & 747.3571 & 0.8350 & 895 \\
\hline
$C_{\rm ll}=1.5$ & 0.1165 & 0.6260 & 1579.9496 & 1.6073 & 983 \\
\hline
$C_{\rm ll}=3$ & 0.1177 & 0.5673 & 947.3134 & 1.2498 & 758 \\
\hline
non-pert scheme (b) & 0.1193 & 0.5923 & 1249.1436 & 1.3957 & 895 \\
\hline
non-pert scheme (c) & 0.1189 & 0.5825 & 1232.5919 & 1.3772 & 895 \\
\hline
%minus non-pert error & 0.1187 & 0.5934 & 841.4468 & 1.3931 & 604 \\
minus non-pert error & 0.1187 & 0.5865 & 1122.1407 & 1.2538 & 895 \\
\hline
%plus non-pert error & 0.1188 & 0.5702 & 939.9997 & 1.5563 & 604 \\
plus non-pert error & 0.1189 & 0.5649 & 1228.4413 & 1.3726 & 895 \\
\hline
\end{tabular}
\caption{\label{tab:CTy3var}
  Central fit result and variations of the results of the CTy3 fit, as described in the text. We indicate in boldface
the central, highest and lowest values.}
\end{center}
\end{table}
These include the scale variations, that we already discussed; the use of different
mass schemes for the definition of the shape variables; variations on the lower limit
of the fit range for each shape variable; three alternative different schemes for the
calculation of the non-perturbative corrections; and the addition or subtraction of the
non-perturbative error to the central value of the non-perturbative correction. In the following
we describe them in turn.
\subsubsection{Mass scheme dependence}
It is easy to realise that there are alternative definitions of the same shape variable that yield the same results
for massless final state particles, but not for massive ones.
For example, in the definitions, the energy and the modulus of the momentum can be used interchangeably when dealing with massless
particles, but this can makes a difference for massive ones. It thus turns out that for the family of schemes
that agree in the massless case, the theoretical result is the same, while the measurements may differ~\cite{Salam:2001bd}. We stress that our calculation of the linear non-perturbative correction
is also insensitive to mass effects, since at the end it always deals with massless particles in the final
state, typically the particles that arise from the splitting of the soft gluon.

In ref.~\cite{Salam:2001bd}, besides the standard schemes commonly used by the
experimental collaborations (reported in Sec.~2 of ref.~\cite{Nason:2023asn}),
three alternative schemes, dubbed E, p and D, where proposed. We assess the effect of the use of different schemes
using a Monte Carlo generator, i.e. we generate a large sample of $e^+e^-\to {\rm hadrons}$ events
using \PythiaEight{}~\cite{Sjostrand:2014zea}, and for each measured distribution, we compute a migration matrix that can be
used to convert from the standard scheme experimental data to the E, p or D scheme. More in detail
for each generated event and for each measured binned event shape distribution, we compute the bin $i$ where the shape variable falls when using the standard scheme, and the bin $j$ where the shape variable falls when using the alternative scheme $S=$E, p or D. A corresponding migration matrix $T^{(S)}_{i,j}$ is then increased by one unit. At the end, the following equation holds
\begin{equation}\label{eq:massmigr}
  n_j^{(S)}=\sum_i n_i^{(\rm Std)} \frac{T^{(S)}_{i,j}}{\sum_k T^{(S)}_{i k}},
\end{equation}
where $n_i^{(S)}$ is the number of events such that the observable computed in the scheme $S$ falls in bin $i$.
We use the right hand side of eq.~\eqref{eq:massmigr} using instead of the $n_i^{(\rm Std)}$ obtained with the
Monte Carlo the one from data, so as to obtain the corrected data for the $S$ scheme.

By inspecting the table, we see that the effect of the mass scheme change yields the largest variation for
the value of $\as$ both in the upper and lower directions. For this reason they are in boldface characters in
the table. Furthermore, we find in all cases very reasonable values of $\chi^2$, so that we do not have a-posteriori reasons to exclude any scheme.

\subsubsection{Fit range}
\label{sec:fitrange}
The computation of the fit range is discussed in section~\ref{sec:fitrange0}, and is controlled by a parameter $C_{ll}$
that is set to two by default. We set $C_{ll}$ to 1.5 and 3 to assess the effect of lowering/rising the lower limit.
We know that our calculation must fail for very low lower limits, due to the raising importance of Sudakov logarithms. We thus
 expect that
the $\chi^2$ should become worse as we lower the lower limit, and be nearly constant as we raise it. This is in fact what
we observe. By raising the limit the change in $\chi^2/{\rm d.o.f.}$ is very small, and the fitted values of
$\alpha(M_Z)$ and $\alpha_0$ change roughly by 0.3\%{} and 4\%{} respectively. On the other hand,
when lowering the limit we get a variation in $\alpha(M_Z)$ and $\alpha_0$ of 1.4\%{} and 6\%{} respectively,
accompanied by a sharp increase in $\chi^2/{\rm d.o.f.}$, warning us not to venture further in that direction.

\subsubsection{Implementation of the non-perturbative correction}
A general summary of the calculation of non-perturbative correction
is given in Sec.~\ref{sec:AppNP}. Following the definitions given there,
besides giving the value in our default scheme, we also present variations
when using schemes (b) and (c).

\subsubsection{Plus or minus the non-perturbative error}
An estimate of missing corrections suppressed by more than  one power
of the non-perturbative scale, we follow the procedure illustrated in Sec.~\ref{sec:AppNP},
that defines the npup and npdn variations.

\subsubsection{General consideration on uncertainties}
The variations reported in table~\ref{tab:CTy3var} are of the order of 1\%.
The most prominent ones are those arising from changing the scheme for the treatment of light hadron masses, being equal to
to +1.6-1.7\%.
Variation of the range lower limit leads also sizeable effect (1.3\%{}). However, in this case we also observe a noticeable increase
in $\chi^2$,  which is not the case when varying the mass schemes.

The problem of ambiguities related to the hadron mass scheme is particularly severe. This is because they are sizeable, but also
because the non-perturbative corrections due to light hadron masses cannot be studied in the large $n_f$ framework, since they always end up
considering final states made up of massless partons. We remark that not many $\as$ fits in the context of $e^+e^-$ shape variables
have considered the mass scheme ambiguities, while we find that they lead to dominant uncertainties.

On the other hand, we find that uncertainties related to how the non-perturbative corrections are included, and to possible higher-order non-perturbative corrections, are small, of the order of half a percent. 

%\commentpn{Make a statement regarding reduced sets of variable}.

\subsection{Comparison to previous parametrization  of non-perturbative effects}
%\subsection{Comparison with results obtained using the previous parametrization
%  of non-perturbative effects}
Older studies of non-perturbative corrections were making use of the large $b_0$ approximate
power corrections evaluate in the 2-jet region extrapolated to the full phase space.
In our framework, this amounts to setting $\zeta(v) \to \zeta(0)$. We have performed such fits
and report the result (including variations) in table~\ref{tab:cty32j}.
\begin{table}[htb]
  \begin{center}
\begin{tabular}{|l|c|c|c|c|c|}
\hline
 Variation & $\as(M_Z)$ & $\alpha_0$ & $\chi^2 $ & $\chi^2/{\rm d.o.f.}$ & d.o.f. \\
\hline
default & 0.1161 & 0.5389 & 1149.9394 & 1.2848 & 895 \\
\hline
$\mu_R=\mu_0/2$ & 0.1155 & 0.6061 & 1523.6604 & 1.8672 & 816 \\
\hline
$\mu_R=2\mu_0$ & 0.1150 & 0.5181 & 1830.6507 & 2.0007 & 915 \\
\hline
std scheme & 0.1153 & 0.4989 & 1106.6396 & 1.2379 & 894 \\
\hline
p scheme & 0.1141 & 0.5119 & 1125.7113 & 1.2592 & 894 \\
\hline
D scheme & 0.1173 & 0.6465 & 923.2022 & 1.0315 & 895 \\
\hline
$C_{\rm ll}=1.5$ & 0.1143 & 0.5658 & 1510.5800 & 1.5367 & 983 \\
\hline
$C_{\rm ll}=3$ & 0.1159 & 0.5325 & 977.2551 & 1.2893 & 758 \\
\hline
non-pert scheme (b) & 0.1163 & 0.5603 & 1281.1125 & 1.4314 & 895 \\
\hline
non-pert scheme (c) & 0.1167 & 0.5305 & 1312.8618 & 1.4669 & 895 \\
\hline
minus non-pert error & 0.1161 & 0.5390 & 1150.0007 & 1.2849 & 895 \\
\hline
plus non-pert error & 0.1161 & 0.5389 & 1149.8783 & 1.2848 & 895 \\
\hline
\end{tabular}
\caption{\label{tab:cty32j} Results for the CTy3 fit using $\zeta(v) \to \zeta(0)$, and
  performing all variations that we have considered.}
\end{center}
\end{table}
We find that in general the $\chi^2$ values that we obtain are as
acceptable as for our standard fits. However, the value of $\as(M_Z)$ that
we obtained are systematically lower by two units in the third digit.

This trend holds also for the fits CT, and C, T, y3
as we can see in table~\ref{tab:allfits2j}.
\begin{table}[htb]
  \begin{center}
    {\small
\begin{tabular}{|l||c|c||c|c||c|c||c|c||}
    \hline
     & \multicolumn{8}{|c|}{$\as(M_Z)$} \\
    \hline
     & \multicolumn{2}{|c||}{CTy3} &  \multicolumn{2}{|c||}{C} &  \multicolumn{2}{|c||}{T} &  \multicolumn{2}{|c||}{$y_3$} \\
    \hline
 Variation & $\zeta(v)$ & $\zeta(0)$ & $\zeta(v)$ & $\zeta(0)$ & $\zeta(v)$ & $\zeta(0)$ & $\zeta(v)$ & $\zeta(0)$ \\
\hline
default & 0.1181 & 0.1161 & 0.1169 & 0.1139 & 0.1168 & 0.1158 & 0.1155 & 0.1154 \\
\hline
$\mu_R=\mu_0/2$ & 0.1167 & 0.1155 & 0.1141 & 0.1105 & 0.1159 & 0.1128 & 0.1122 & 0.1131 \\
\hline
$\mu_R=2\mu_0$ & 0.1167 & 0.1150 & 0.1212 & 0.1184 & 0.1208 & 0.1191 & 0.1157 & 0.1161 \\
\hline
std scheme & 0.1173 & 0.1153 & 0.1164 & 0.1118 & 0.1152 & 0.1148 & 0.1150 & 0.1149 \\
\hline
p scheme & 0.1160 & 0.1141 & 0.1164 & 0.1118 & 0.1152 & 0.1148 & 0.1137 & 0.1135 \\
\hline
D scheme & 0.1199 & 0.1173 & 0.1190 & 0.1153 & 0.1205 & 0.1170 & 0.1168 & 0.1166 \\
\hline
$C_{\rm ll}=1.5$ & 0.1165 & 0.1143 & 0.1151 & 0.1116 & 0.1154 & 0.1133 & 0.1142 & 0.1142 \\
\hline
$C_{\rm ll}=3$ & 0.1177 & 0.1159 & 0.1221 & 0.1116 & 0.1180 & 0.1172 & 0.1156 & 0.1154 \\
\hline
non-pert scheme (b) & 0.1193 & 0.1163 & 0.1191 & 0.1176 & 0.1185 & 0.1184 & 0.1154 & 0.1154 \\
\hline
non-pert scheme (c) & 0.1189 & 0.1167 & 0.1195 & 0.1172 & 0.1192 & 0.1191 & 0.1154 & 0.1154 \\
\hline
minus non-pert error & 0.1187 & 0.1161 & 0.1173 & 0.1139 & 0.1165 & 0.1158 & 0.1157 & 0.1154 \\
\hline
plus non-pert error & 0.1189 & 0.1161 & 0.1172 & 0.1139 & 0.1172 & 0.1158 & 0.1153 & 0.1154 \\
\hline
\end{tabular}
}
\end{center}
\caption{\label{tab:allfits2j}
Summary of $\alpha_s(M_Z)$ results for all fits and variations considered.}
\end{table}

\subsection{Including the heavy-jet mass and the jet-mass difference in the fits}
\label{sec:masses}
\begin{figure}[htb]
  \begin{center}
    \includegraphics[width=0.32\linewidth, page=1]{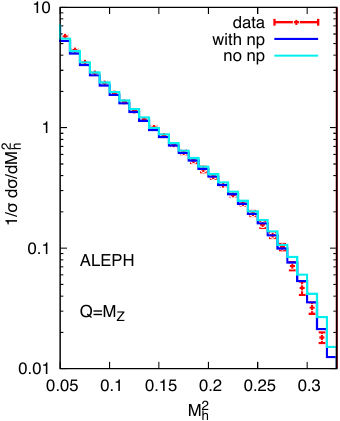}
    \includegraphics[width=0.32\linewidth, page=2]{Mhd-plot/Mh2Md2All-crop.pdf}
    \includegraphics[width=0.32\linewidth, page=3]{Mhd-plot/Mh2Md2All-crop.pdf}    
    \caption{Comparison of ALEPH, DELPHI and WICKE data for $M_h^2$ distribution at $Q = M_Z$. }
    \label{fig:mh2}
  \end{center}
\end{figure}

\begin{figure}[htb]
  \begin{center}
    \includegraphics[width=0.32\linewidth, page=4]{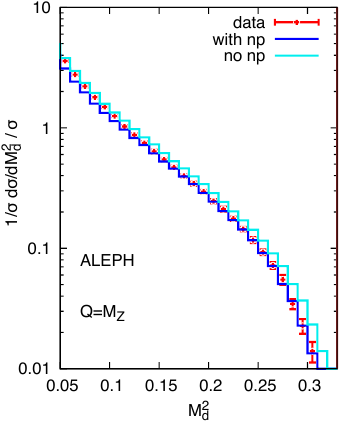}
    \includegraphics[width=0.32\linewidth, page=5]{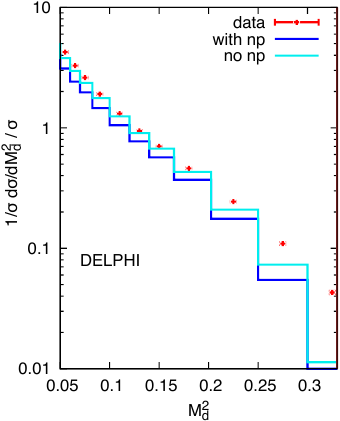}
    \includegraphics[width=0.32\linewidth, page=6]{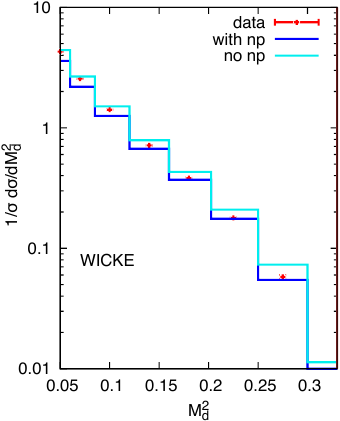}    
    \caption{Same as Fig.~\ref{fig:mh2} for the $M_d^2$ distribution.}
    \label{fig:md2}
  \end{center}
\end{figure}

In ref.~\cite{Nason:2023asn} it was shown that the $\zeta$ functions for the heavy jet mass
and for the jet mass difference undergo very sharp variations near the two jet limit configuration.
We then preferred not to include them in our fit. Nevertheless, we were able to show that
their behaviour in the three-jet region is in nice agreement with our calculation of non-perturbative
corrections, that for these observables are in fact negative.
It is thus interesting to see what happens if one attempts to include also
these observables in the fits.
In order to do this, we face an immediate problem, namely that the
DELPHI data seem to be inconsistent with data from other
experiments.
Since the ALEPH experiment provides very precise data, which are in
agreement with all other experiments, in the following, we show
comparisons of DELPHI to ALEPH data at the $Z$-peak as an
illustration.  Since the two experiments use a very different binning,
to illustrate the comparison, it is useful to take as a reference the
theoretical prediction, which we show both including and omitting the
non-perturbative correction. This comparison is displayed in
Fig.~\ref{fig:mh2} and~\ref{fig:md2} left and central panels.
It is clear that ALEPH data, which use a much finer binning than the
DELPHI ones, are in excellent agreement with the theory prediction
which includes the non-perturbative correction. Instead, the DELPHI
data is considerably above both predictions. This is the case for both
mass distributions. This is a clear indication that the experimental
data are not consistent with each other. On the other hand, DELPHI
data have been also re-analysed at $Q=M_Z$ in
ref.~\cite{Wicke:1999zz}. These data, that we label ``WICKE'' in the following, are shown in
the right panels of Figs.~\ref{fig:mh2} and~\ref{fig:md2}. We notice
that these data are in good agreement with theory predictions
including non-perturbative corrections, and therefore are compatible
with ALEPH (and all other) data. For this reason, in the following, we
have decided to omit in the fit all DELPHI data on $\rm M_h^2$ and
$\rm M_d^2$ and to include instead data from ref.~\cite{Wicke:1999zz},
that we denote as ``WICKE''.\footnote{We found no evidence of other
  relevant discrepancies in the data for other shape variables, in
  particular also in the DELPHI data, that are thus kept in the fit.}

In Table~\ref{tab:datamasses} we show all additional data that we have
included in the fit. We use as upper limits for the fit ranges 0.33 in both
cases, while the lower limits are set as described before.
In Table~\ref{tab:fitwithmasses} we show the results of our fit. 
\begin{table}[htb]
\begin{center}
\begin{tabular}{|l|l|l|}
\hline Obs. & Experiment & Energies in GeV \\ \hline\hline $M_d^2
$&ALEPH~\cite{ALEPH:2003obs} & 91.2 133 161 172 183 189 200 206
\\ \hline $M_h^2$& WICKE~\cite{Wicke:1999zz} & 91.2 \\ \hline $M_h^2
$&ALEPH~\cite{ALEPH:2003obs} & 91.2 133 161 172 183 189 200 206
\\ \hline $M_h^2$&JADE~\cite{MovillaFernandez:1997fr} & 35 44
\\ \hline $M_h^2$&L3~\cite{L3:1992nwf,L3:2002oql,L3:2004cdh} & 41.4
55.3 65.4 75.7 82.3 85.1 91.2 130.1 136.1 161.3 172.3 \\ & & 182.8
188.6 194.4 200 \\ \hline
$M_h^2$&OPAL~\cite{OPAL:1992dnu,OPAL:1993pnw,OPAL:2004wof} & 91.2 133
177 197 \\ \hline $M_h^2$&SLD~\cite{SLD:1994idb} & 91.2 \\ \hline
$M_h^2$&TRISTAN~\cite{TOPAZ:1993vyh} & 58 \\ \hline $M_h^2$&
WICKE~\cite{Wicke:1999zz} & 91.2 \\ \hline
\end{tabular}
\end{center}
\caption{\label{tab:datamasses}
Summary of data for $M_h^2$ and $M_d^2$ used in the section.}
\end{table}
\begin{table}[htb]
\begin{center}
\begin{tabular}{|l|c|c|c|c|c|}
\hline
 Variation & $\as(M_Z)$ & $\alpha_0$ & $\chi^2 $ & $\chi^2/{\rm d.o.f.}$ & d.o.f. \\
\hline
default & {\bf 0.1214} & 0.5214 & 2737.0665 & 2.1267 & 1287 \\
\hline
$\mu_R=\mu_0/2$ & 0.1200 & 0.5140 & 4602.1494 & 3.8968 & 1181 \\
\hline
$\mu_R=2\mu_0$ & 0.1205 & 0.5380 & 4056.9046 & 3.1040 & 1307 \\
\hline
std scheme & 0.1221 & 0.4649 & 6032.3920 & 4.7055 & 1282 \\
\hline
p scheme & {\bf 0.1189} & 0.5007 & 2540.5896 & 1.9756 & 1286 \\
\hline
D scheme & {\bf 0.1253} & 0.6041 & 3605.6129 & 2.8016 & 1287 \\
\hline
$C_{\rm ll}=1.5$ & 0.1224 & 0.5252 & 4238.7596 & 3.0191 & 1404 \\
\hline
$C_{\rm ll}=3$ & 0.1199 & 0.5239 & 2076.6755 & 1.8692 & 1111 \\
\hline
%non-pert scheme (b) & Warning:alphasatlowestpointinrange \\
non-pert scheme (b) & 0.1220   &  0.4810   &  4494.4234  &   3.4922 & 1287 \\
\hline
non-pert scheme (c) & 0.1221 & 0.4914 & 4148.3055 & 3.2232 & 1287 \\
\hline
minus non-pert error & 0.1216 & 0.5186 & 2769.4522 & 2.1519 & 1287 \\
\hline
plus non-pert error & 0.1212 & 0.5243 & 2710.8829 & 2.1064 & 1287 \\
\hline
\end{tabular}
\end{center}
\caption{\label{tab:fitwithmasses} As Table~\ref{tab:CTy3var} but
  including also the data for $M_h^2$ and $M_d^2$ listed in
  Table~\ref{tab:datamasses}.}
\end{table}
We find that the inclusion of the masses in the fits increases the
central value of $\as$ by about 2.8\% and worsens in general all fits,
yielding considerably larger $\chi^2$/d.o.f. values. Both features are
mitigated when raising the lower limit of the fit range. This finding
  supports the view that perhaps the sharp variation of the $\zeta$ functions
  near the two-jet limit, found in ref.~\cite{Nason:2023asn}, introduces further uncertainty
  in the result near the two-jet region.

In Table~\ref{tab:fitwithmasses2J} we show the same fit including
mass-distributions, but now using the non-perturbative corrections
computed in the two-jet limit. We find a reduction in the value of the
fitted $\as(M_Z)$, and considerably larger $\chi^2$ values,
inidicating that constant non-perturbative corrections are
incompatible with data.
\begin{table}[htb]
  \begin{center}
    \begin{tabular}{|l|c|c|c|c|c|}
\hline
 Variation & $\as(M_Z)$ & $\alpha_0$ & $\chi^2 $ & $\chi^2/{\rm d.o.f.}$ & d.o.f. \\
\hline
default & 0.1145 & 0.5769 & 6576.4898 & 5.1099 & 1287 \\
\hline
$\mu_R=\mu_0/2$ & 0.1163 & 0.5983 & 3801.7747 & 3.2191 & 1181 \\
\hline
$\mu_R=2\mu_0$ & 0.1068 & 0.6053 & 11697.8542 & 8.9502 & 1307 \\
\hline
std scheme & 0.1174 & 0.5225 & 5616.3416 & 4.3809 & 1282 \\
\hline
p scheme & 0.1110 & 0.5633 & 6437.6362 & 5.0059 & 1286 \\
\hline
D scheme & 0.1158 & 0.6806 & 10599.5174 & 8.2358 & 1287 \\
\hline
$C_{\rm ll}=1.5$ & 0.1138 & 0.5865 & 6971.5191 & 4.9655 & 1404 \\
\hline
$C_{\rm ll}=3$ & 0.1159 & 0.5735 & 5808.1213 & 5.2278 & 1111 \\
\hline
non-pert scheme (b) & 0.1170 & 0.5693 & 8415.0241 & 6.5385 & 1287 \\
\hline
non-pert scheme (c) & 0.1167 & 0.5491 & 8300.0328 & 6.4491 & 1287 \\
\hline
minus non-pert error & 0.1145 & 0.5770 & 6576.7989 & 5.1102 & 1287 \\
\hline
plus non-pert error & 0.1145 & 0.5769 & 6576.1813 & 5.1097 & 1287 \\
\hline
\end{tabular}
\end{center}
\caption{\label{tab:fitwithmasses2J} As Tables~\ref{tab:fitwithmasses} but with power corrections computed in the two-jet limit.}
\end{table}

We stress that, for the mass distributions, we have not included
resummation effects in the two-jet limit, consistenly with what is
done in the rest of the paper, nor all-order effects close to the
symmetric three-particle limit ($M_h^2 = M_d^2 = 1/3$), as studied in
ref.~\cite{Bhattacharya:2023qet}.
We find however that lowering the upper limit in our fit does not
change the results in a noticeable way.

\section{Conclusions}
\label{sec:conclu}

In this work, we have presented an extension of an analysis carried
out in ref.~\cite{Nason:2023asn}, where we have fitted $e^+e^-$
shape-variable data using newly found results on power-corrections in
the three-jet region~\cite{Caola:2022vea}. While in
ref.~\cite{Nason:2023asn} only data collected at the $Z$ peak were
considered, here we consider instead a large set of experiments and
energies. We find results that are consistent with
ref.~\cite{Nason:2023asn}, characterized by a value of $\as(M_Z)$ well
compatible with the world average but with sizeable
uncertainties. Consistently with what observed in
ref.~\cite{Nason:2023asn} also here we find that the dominant
uncertainty is to due the ambiguity in the mass scheme.

A nice feature of this new analysis is that now, unlike in
ref.~\cite{Nason:2023asn}, if we fit each variable
individually, we also find reasonable determinations of the strong
coupling, since the availability of different energies allows one to
disentangle perturbative and non-perturbative effects. However, the
associated errors for the individual fits are much larger than the
one for the combined fit.

One feature of the newly found three-jet power corrections is that
they are negative for the heavy-jet mass and jet-mass
difference. Experimental data support this finding, as already
remarked in ref.~\cite{Nason:2023asn}. Here we have shown that if we
include the heavy-jet mass and jet-mass difference in the fits, in general
we obtain large $\chi^2$ values, that are however much worse if we
use as non-perturbative input the 2-jet determinations extrapolated
to the three-jet region, as was done in earlier works.

In general, we stress that variations in the theory predictions lead
to values of $\as$ that differ by up to three percent. This suggests
that, once all relevant uncertainties are taken into account, a fit of
$\as$ with an uncertainty below a percent seems out of reach.

We emphasize that our implementation of non-perturbative effects is the
minimal and most straighforward one to verify the impact of
non-perturbative corrections away from the two-jet limit.
A further development of our work could be the inclusion of
resummmation effects, together with a better treatment of the
interplay between soft radiation and non-perturbative corrections.

\bibliographystyle{JHEP}
\bibliography{Shapes.bib}
 
\end{document}